\documentclass[prd, nofootinbib, floatfix, notitlepage]{revtex4-1}

\usepackage{amsmath,amsfonts,amssymb}
\usepackage{mathrsfs}
\usepackage{graphicx}
\usepackage[english]{babel} 
\usepackage{hyperref} 
 \usepackage[normalem]{ulem}
   \usepackage{textcomp}

\usepackage{color}
\usepackage{bbold}

\usepackage{hyperref}
 \usepackage[normalem]{ulem}
   \usepackage{textcomp}
   \usepackage{xcolor}

\begin{document}

\title{LISA for Cosmologists: Calculating the Signal-to-Noise Ratio for Stochastic and Deterministic Sources}
\author{Tristan L.~Smith${}^1$ and Robert R.~Caldwell${}^2$}

\affiliation{$^1$Department of Physics \& Astronomy, Swarthmore College, Swarthmore, PA 19081 USA\\ $^2$Department of Physics \& Astronomy, Dartmouth College, Hanover, NH 03755 USA}

\date{\today}

\begin{abstract}
We present the steps to forecast the sensitivity of the Laser Interferometer Space Antenna (LISA) to both a stochastic gravitational wave background and deterministic wave sources. We show how to use these expressions to estimate the precision with which LISA can determine parameters associated with these sources. Tools are included to enable easy calculation of the signal-to-noise ratio and draw sensitivity curves. Benchmark values are given for easy comparison and checking of methods in the case of three worked examples. The first benchmark is the threshold stochastic gravitational wave background $\Omega_{GW} h^2$ that LISA can observe. The second is the signal-to-noise ratio that LISA would observe for a binary black hole system identical to GW150914, radiating four years before merger. The third is the case of a monotone source, such as a binary that is far from merger.
\end{abstract}

\maketitle

\section{Explanation}

This document is intended to be used as a set of instructions for calculating the sensitivity of the Laser Interferometer Space Antenna (LISA) \cite{Audley:2017drz} to a stochastic gravitational wave background (SGWB) or a continuous wave source under idealized circumstances. By idealized we specifically mean that all noise is Gaussian, stationary, and that there are no foregrounds. Many of the results presented are well known, and have a long presence in the literature (e.g. Refs.~\cite{Allen:1997ad,Maggiore:1999vm,Cornish:2001qi,Cornish:2001bb,Moore:2014lga,Thrane:2013oya,Romano:2016dpx}). However, we perceive that an accessible introduction is lacking. Moreover, we are unaware of any literature that gives a complete, end-to-end derivation of the signal-to-noise ratio for LISA in its current design. Our goal is to facilitate sensitivity calculations, in the hope that more theorists will be able to properly evaluate the ability of LISA to detect and distinguish their favorite sources.  We have tried to write the type of document that we wish we had when we started our investigations. In order that these tools are not just a black box, we have included some basic derivations which allows for an extension to other interferometric designs. The calculation of the sensitivity presented here gives a straight-forward accounting for the standard time-delay interferometry (TDI) signals and explains how the monitoring of the instrumental noise using the Sagnac signal leads to a significant increase in sensitivity. For the impatient reader who wants to evaluate the signal-to-noise ratio or forecast parameter sensitivity for a stochastic background, here are the key results.

The signal-to-noise ratio (SNR) for a SGWB is given in Eqs.~(\ref{eqn:altSNR}) and (\ref{eqn:box2}) as
\begin{equation}
\boxed{
    {\rm SNR}^2 = T \int_0^\infty df \, \frac{\Omega_{GW}^2}{\Sigma_\Omega^2}, \quad
    \Sigma_\Omega = \Sigma_I \frac{4 \pi^2 f^3}{3 H_0^2}, \quad 
    \Sigma_I \simeq \sqrt{2} \frac{20}{3} \left[ \frac{S_I(f)}{(2 \pi f)^4} + S_{II}(f) \right]\left[1+\left(\frac{f}{ 4f_*/3}\right)^2\right],
    }
    \nonumber
\end{equation}
where $\Sigma_\Omega$, $\Sigma_I$ are the inverse-noise weighted sensitivity to the spectral density and intensity for two TDI modes, $T$ is the observation time, where the nominal mission lifetime is $4$ years, $f_* = c/(2 \pi L)$, $L=2.5\times 10^6$~km, and $S_I,\, S_{II}$ are given in Eqs.~(\ref{eqn:SI}) and (\ref{eqn:SII}). The last expression is made under a low-frequency assumption. A worked example is provided in  Sec.~\ref{sec:integrated}.

The signal-to-noise ratio for a deterministic source such as an inspiraling binary is given in Eqs.~(\ref{eqn:LISASNR}) and (\ref{eqn:scird}) as
\begin{equation}
\boxed{
    {\rm SNR}^2 = \int_0^\infty df \, \frac{\bar{h}^2(f)}{\Sigma_{h}(f)}, \quad  
    \Sigma_{h}(f) \simeq \frac{1}{2} \, \frac{20}{3} \, \left[ \frac{S_I(f)}{(2 \pi f)^4} + S_{II}(f) \right] R(f),
    }
\nonumber
\end{equation}
where $\Sigma_h$ is the inverse-variance weighted waveform sensitivity for two TDI modes, $\bar h$ is the sky-, polarization-, and orientation-averaged waveform amplitude as defined in Eq.~(\ref{eqn:hwaveform}), and $R(f) = 1 + (f/f_2)^2$ and $f_2 = 25$~mHz. The last expression is again made under a low-frequency assumption. A worked example is provided in Sec.~\ref{sec:binary}.

The layout of the article is as follows. In Sec.~\ref{sec:GW} we introduce our notation. In Sec.~\ref{sec:signal} we introduce the form of the signal and noise for the TDI modes, and calculate the detector response. In Sec.~\ref{sec:optimal1} we present the calculation of the optimal statistic for a stochastic background, and in Sec.~\ref{sec:LISAnoise} we introduce the LISA noise model. In Sec.~\ref{sec:skyavg} we present the calculation of the optimal statistic for a sky- and polarization-averaged deterministic point source. Three examples are presented in Sec.~\ref{sec:examples}. We wrap up in Sec.~\ref{sec:discuss}. Table \ref{tab:notation} lists the variables used in this paper. Finally, we also provide a Mathematica notebook to enable easy calculations.\footnote{\url{https://doi.org/10.5281/zenodo.3341817}}

\section{Gravitational Waves}
\label{sec:GW}

We begin by establishing our notation and conventions. We expand the gravitational-wave metric perturbation in plane waves with respect to a coordinate system at rest relative to the Solar System barycenter:
\begin{equation}
h_{ab}(\vec x,t) = \int_{-\infty}^{\infty}df  \int d^2\hat n \, \sum_P \,  h_P(f,\hat n) e^P_{ab}(\hat{n})e^{i2 \pi f(t - \hat n \cdot \vec x/c)},\label{eq:Fourier}
\end{equation}
where $e^P_{ab}$ is the polarization tensor.  For a $P=+,\,\times$ polarized plane wave propagating in the $\hat n$ direction, the polarization tensors may be written
\begin{eqnarray}
\hat n &=& (\cos\phi \, \sin\theta, \sin\phi\, \sin\theta, \cos\theta) \\ 
e^{+}_{ab}(\hat n) &=& \hat{ \mathbb{m}}_a \hat{\mathbb{m}}_b - \hat{\mathbb{n}}_a \hat{\mathbb{n}}_b \label{eq:epab}\\
e^{\times}_{ab}(\hat n) &=& \hat{ \mathbb{m}}_a \hat{ \mathbb{n}}_b + \hat{ \mathbb{n}}_a\hat{ \mathbb{m}}_b \label{eq:ecab}\\
\hat{ \mathbb{m}} &\equiv& (\sin\phi, -\cos\phi, 0) \\
\hat{ \mathbb{n}} &\equiv& (\cos\phi\, \cos\theta, \sin\phi\, \cos\theta, -\sin\theta)
\end{eqnarray}
so that $e^{P}_{ab}(\hat n) e^{P'ab}(\hat n) = 2 \delta_{P P'}$ and $\hat{ \mathbb{m}},\,\hat{ \mathbb{n}}$ are basis (Newman-Penrose) vectors that define the coordinate system in the plane transverse to the direction of propagation. 

We assume that the SGWB is Gaussian-distributed and has zero-mean so that its properties are characterized in terms of the variance or power spectrum (i.e., the spectral density). Considering the possible polarization states, we express the covariance in terms of Stokes parameters, as
\begin{equation}
\left(\begin{array}{cc}\langle h_+^*(f,\hat{n}) h_+(f',\hat{n}')\rangle & \langle h_+^*(f,\hat{n}) h_{\times}(f',\hat{n}')\rangle \\ \langle h_\times^*(f,\hat{n}) h_+(f',\hat{n}')\rangle & \langle h_\times^*(f,\hat{n}) h_{\times}(f',\hat{n}')\rangle\end{array}\right) = \frac{1}{2} \delta_D(f-f') \frac{\delta^{(2)}(\hat{n} - \hat{n}')}{4\pi} \left(\begin{array}{cc}I +Q & U+iV \\U-iV & I-Q\end{array}\right). \label{eq:stokes}
\end{equation}
The overall intensity, $I$, and circular polarization, $V$, are scalar quantities, and hence can be measured through the monopole of the stochastic background; the $Q$ and $U$ are spin-4 quantities and hence do not contribute to an isotropic, stochastic, background. Since we are considering the intensity of an isotropic background, for the rest of this discussion we will take $V=Q=U=0$. Note that the intensity is related to the spectral density of the SGWB,
\begin{equation}
\Omega_{\rm GW} \equiv \frac{d \ln \rho_{GW}}{d\ln f} = \left(\frac{4\pi^2}{3H_0^2}\right) f^3 I(f).\label{eq:Omega_gw}
\end{equation}
Our notation agrees with Refs.~\cite{Maggiore:1999vm,Cornish:2001bb}, where a signal power is defined such that $S_h(f) = I(f)$. We caution that in Refs.~\cite{Thrane:2013oya,Moore:2014lga,Romano:2016dpx}, a signal power $S_h = 2 I$ is defined; this alternate convention is offset by another factor of two, elsewhere in those references.

\section{The Signal and Covariance}
\label{sec:signal}

\begin{figure}[b]
\begin{center}
\resizebox{!}{5cm}{\includegraphics{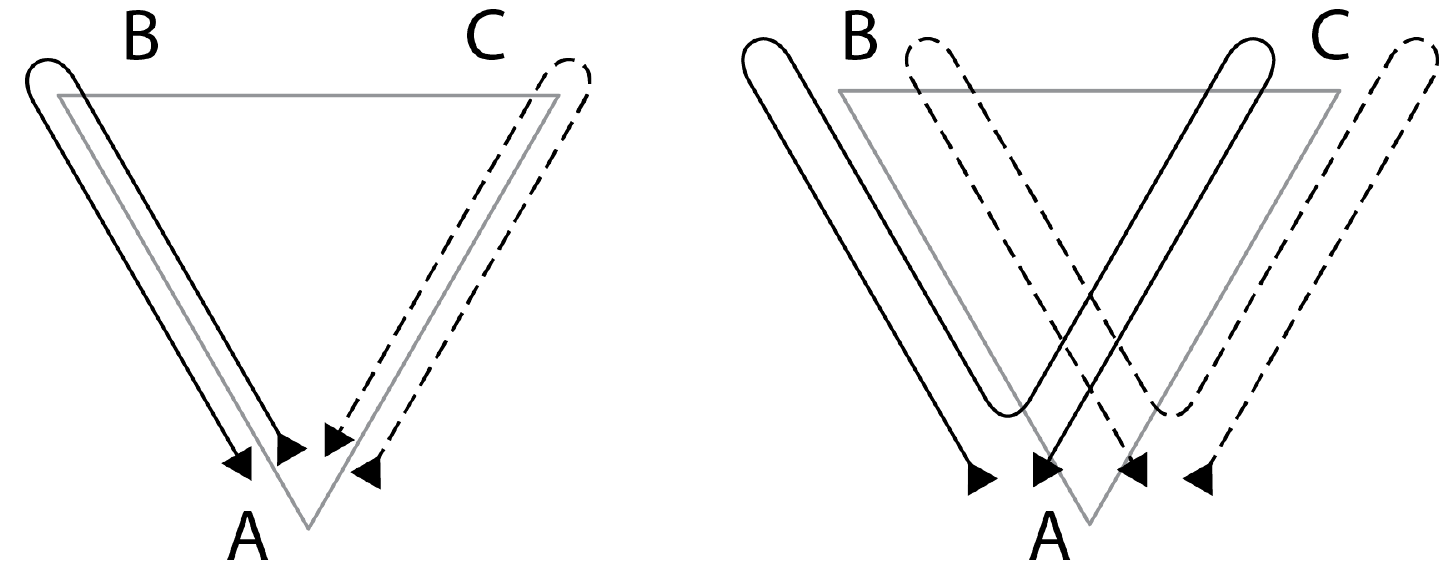}}
\caption{(Left) The two round-trip interferometer paths used to compose the Michelson signal. (Right) The two pairs of round-trip interferometer paths used to compose the Michelson TDI signal $\Delta\varphi_{A_{BC}}$.}
\label{fig:ThreeSixLink}
\end{center}
\end{figure}

 The measured phase difference at a vertex of the interferometer, $\Phi$, can be written in terms of the gravitational response in terms of an interferometer phase at that vertex, $\Delta\varphi$, as well as the noise, $n$,  
\begin{equation}
    \Phi_{A_{BC}}(t) = \Delta\varphi_{A_{BC}}(t) + n_{A_{BC}}(t).
\end{equation}
The subscript $A_{BC}$ indicates the signal at the interferometer consisting of arms $AB$ and $AC$, as shown in Fig.~\ref{fig:ThreeSixLink}. 
It is straightforward but tedious to show that the phase difference measured at that vertex is given by \cite{Cornish:2001bb}
\begin{equation}
\Delta\varphi_{A_{BC}}(t) = \int_{-\infty}^\infty df \int d^2 \hat n \sum_P {h}_P(f,\hat n)e^{i 2\pi f t} F^P_{A_{BC}}(\hat n,f;t),\label{eq:int_phase}
\end{equation}
where
\begin{eqnarray}
F^P_{A_{BC}}(\hat n,f;t) &= & \frac{1}{2} e^{-i 2\pi f\hat{n} \cdot \vec x_A(t)/c}e_{ab}^P(\hat n) \left[ \mathcal{F}^{ab}(\hat{\ell}_{AB}(t) \cdot \hat{n},f) -\mathcal{F}^{ab}(\hat{\ell}_{AC}(t) \cdot \hat{n},f)\right] \label{eq:gain1} \\
\mathcal{F}^{ab}(\hat{\ell} \cdot \hat n,f) &=&
\frac{1}{2}W(f,f_*)\hat{\ell}^a \hat{\ell}^b \left( {\rm sinc}\left[\frac{f}{2 f_*}(1-\hat{\ell}\cdot\hat n)\right] e^{-i\frac{f}{2f_*}(3+\hat{\ell}\cdot\hat n)} + {\rm sinc}\left[\frac{f}{2 f_*}(1+\hat{\ell}\cdot\hat n)\right] e^{-i\frac{f}{2f_*}(1+\hat{\ell}\cdot\hat n)} \right)\label{eq:gain2}
\end{eqnarray}
gives the gain of a detector vertex. The vector $\hat \ell_{AB}$ points from vertex $A$ to $B$, $f_* = c/(2 \pi L)$, and $\hat n$ is the direction of gravitational wave propagation. When $W=1$, the above expressions fully account for the round trip paths $ABA$ and $ACA$ as illustrated in the left panel of Fig.~\ref{fig:ThreeSixLink}. To mitigate additional sources of noise, TDI uses longer paths, which are illustrated in the right panel of Fig.~\ref{fig:ThreeSixLink}. The phase accumulated by these additional paths are given by the same expressions, but with the time offset by a factor $t \to t-2 L/c$. The offset in time results in a phase shift in the time-series Fourier transform. Hence, the factor $W(f,f_*) =1-e^{-2 i f/f_*}$ accounts for the full round trips of the TDI signal. 

To detect the irreducible hum of a SGWB we correlate the response between the different vertices of the constellation of detectors. Assuming that the SGWB and the noise are uncorrelated, the full response is a sum of the gravitational wave signal and noise: 
\begin{eqnarray}
\langle \Phi_{A_{BC}}(t) \, \Phi_{X_{YZ}}(t') \rangle &=& \frac{1}{2} \int_{-\infty}^\infty df e^{i 2\pi f(t-t')}\left[ \mathcal{R}_{A_{BC},X_{YZ}}(f;t,t') I(f)  + N_{A_{BC},X_{YZ}}(f) \right],
\label{eqn:rdefn}
\end{eqnarray}
where the SGWB intensity response function $\mathcal{R}$ for a given detector geometry is given by 
\begin{equation}
\mathcal{R}_{A_{BC},X_{YZ}}(f;t,t') = \int \frac{d^2 \hat{n}}{4\pi}\left[F^+_{A_{BC}}(\hat n,f;t)F^{*+}_{X_{YZ}}(\hat n,f;t')+F^\times_{A_{BC}}(\hat n,f;t)F^{*\times}_{X_{YZ}}(\hat n,f;t')\right],
\label{eqn:rint}
\end{equation}
and $N_{A_{BC},X_{YZ}}(f)$ is the correlated noise power between the two detectors. Note that the response depends on time because of the orbital motion of the spacecraft, although we will ultimately ignore this feature. When assessing the sensitivity of LISA to a SGWB we wish to determine the minimum intensity $I$ that can be determined in the presence of noise $N$, as a function of frequency. 

Specializing to LISA (with three spacecraft arranged on the vertices of a fixed equilateral triangle) we can write the full covariance for the phases measured at each spacecraft as 
\begin{eqnarray}
\langle \Phi_J(t){\Phi}_{J'}(t')\rangle = \left(\begin{array}{ccc}C_1  & C_2   & C_2   \\ C_2   & C_1   & C_2  \\ C_2  & C_2   & C_1  \end{array}\right),\label{eq:noise1}
\end{eqnarray}
where $J,\,J' = \{A_{BC},\,B_{CA},\,C_{AB}\}$. Elsewhere in the literature, $A_{BC},\,B_{CA},\,C_{AB}$ are labeled as $X,\,Y,\, Z$ \cite{Vallisneri:2012np}. The correlations $C_{1,2}$ consist of a contribution from the SGWB and from the instrument noise, {\it i.e.} 
\begin{equation}
    C_i = {\cal S}_i + N_i,
\end{equation} 
where ${\cal S}$ is the signal power convolved with the instrument gain, and $N$ is the instrument noise power. 
  
We can construct three orthogonal (i.e., statistically independent) signals by diagonalizing the above covariance matrix. Note that by diagonalizing the covariance, the cross-correlation between the TDI variables has zero response \emph{to both the instrumental noise and the SGWB.} The  response eigenvectors are
\begin{eqnarray}
\Phi_I &=& \frac{1}{\sqrt{6}}\left(\Phi_{A_{BC}} - 2\Phi_{B_{CA}}+\Phi_{C_{AB}}\right)\\
\Phi_{II} &=& \frac{1}{\sqrt{2}}\left(\Phi_{A_{BC}} - \Phi_{C_{AB}}\right)\\
\Phi_{III} &=& \frac{1}{\sqrt{3}}\left(\Phi_{A_{BC}} + \Phi_{B_{CA}} + \Phi_{C_{AB}}\right)
\end{eqnarray}
which yield a diagonal covariance matrix with entries
\begin{eqnarray}
C_I = C_{II} &=& C_1 - C_2 \label{eqn:IandII}\\
C_{III} &=& C_1 + 2 C_2.\label{eqn:III}
\end{eqnarray}
The eigenmodes $I,\,II,\,III$ are none other than the time-delay interferometry (TDI) variables A, E, and T. For the rest of this document we will refer to these eigenmodes by their TDI labels \cite{1999ApJ...527..814A,Tinto:2002de,Prince:2002hp}. Hence,
\begin{eqnarray}
N_{\rm A} = N_{\rm E} &=& N_1-N_2 \label{eq:tdinoise1}\\
N_{\rm T} &=& N_1 + 2 N_2\label{eq:tdinoise2} \\
 {\cal S}_{\rm A} = {\cal S}_{\rm E} &=&   {\cal S}_1-{\cal S}_2 \label{eq:tdisig1} \\
 {\cal S}_{\rm T} &=&  {\cal S}_1+2 {\cal S}_2.\label{eq:tdisig2}
\end{eqnarray}
(For comparison, see Eqs.~(19-22) of Ref.~\cite{Vallisneri:2012np}.) We note that the autocorrelations ${\cal S}_1$ and $N_1$ are sometimes referred to as the Michelson signal and noise \cite{Cornish:2001bb,Thrane:2013oya}. Confusingly, the eigenmodes A, E, T with $W=1$ are also sometimes referred to as Michelson modes. It might make more sense if they were called Michelson {\it eigen}-modes. Finally, we refer to the eigenmodes A, E, and T  with $W = 1-e^{-2 i f/f_*}$ as TDI modes.

The intensity response function in Eq.~(\ref{eqn:rint}) has the property that in the limit of vanishing frequency, for a single Michelson interferometer,
\begin{equation}
{\cal R}_{A_{BC},A_{BC}} \to \frac{2}{5} \sin^2\beta 
\end{equation}
where $\beta=\pi/3$ is the angle between the detector arms for LISA. We use the same convention as Refs.~\cite{Cornish:2001bb,Maggiore:1999vm}. On the other hand, the response function denoted $\Gamma$ in Refs.~\cite{Thrane:2013oya,Moore:2014lga,Romano:2016dpx} is related to ours by a factor ${\cal R} = 2\Gamma$. This factor of two compensates for the different, previously mentioned factor in the signal power.

We specify the response of LISA to gravitational waves as follows. For this simplified discussion, we model the position of the spacecraft as fixed in space,
\begin{eqnarray}
\vec x_A &=& \{0,0,0\} \\
\vec x_B &=& L \{1/2, \sqrt{3}/2, 0 \} \label{eq:length}\\
\vec x_C &=& L \{-1/2, \sqrt{3}/2, 0\}
\end{eqnarray}
where $L = 2.5\times 10^9$~m, or $25/3$~cs (light-seconds). Even though the spacecraft are moving relative to the solar system barycenter (the frame in which we expand the gravitational wave signal), the optimal statistic effectively filters any correlation with a time-lag much greater than the light-travel time across the constellation \cite{Cornish:2001bb}. In this discussion we ignore the relative motion between the instantaneous frames of the spacecraft within this time-lag. See Ref.~\cite{Cornish:2003tz} for more details.

The intensity response functions  for the SGWB covering the full frequency range must be calculated numerically.  We can obtain an analytic approximation by expanding the gain of a detector vertex [given in Eqs.~(\ref{eq:gain1}) and (\ref{eq:gain2})] in powers of $x \equiv f/f_* \ll 1$ and integrating that expansion over the sky to obtain:
\begin{eqnarray}
\mathcal{R}_1(f;t,t) &\simeq& 
|W|^2\left( \frac{3}{10}
-\frac{169}{1680}x^2 
+\frac{85}{6048}x^4 
-\frac{165073}{159667200}x^6 
+\frac{132439}{2830464000}x^8
+ {\cal O}(x^{10})\right),
\label{eqn:R1}\\
\mathcal{R}_2(f;t,t) &\simeq&
|W|^2\left(-\frac{3}{20}
+\frac{169}{3360}x^2 
-\frac{85}{12096}x^4 
+\frac{29239}{45619200}x^6 
-\frac{251389}{5660928000}x^8
+ {\cal O}(x^{10})\right),
\label{eqn:R2}
\end{eqnarray}
where we have left the factor $|W|^2$ intact to enable switching between Michelson and TDI variables. We have expanded these response functions to such a high power so that we keep the leading and next-to-leading orders in the expansion for the T response $\mathcal{S}_1+2\mathcal{S}_2$.

\begin{figure}[t]
\begin{center}
\resizebox{!}{7cm}{\includegraphics{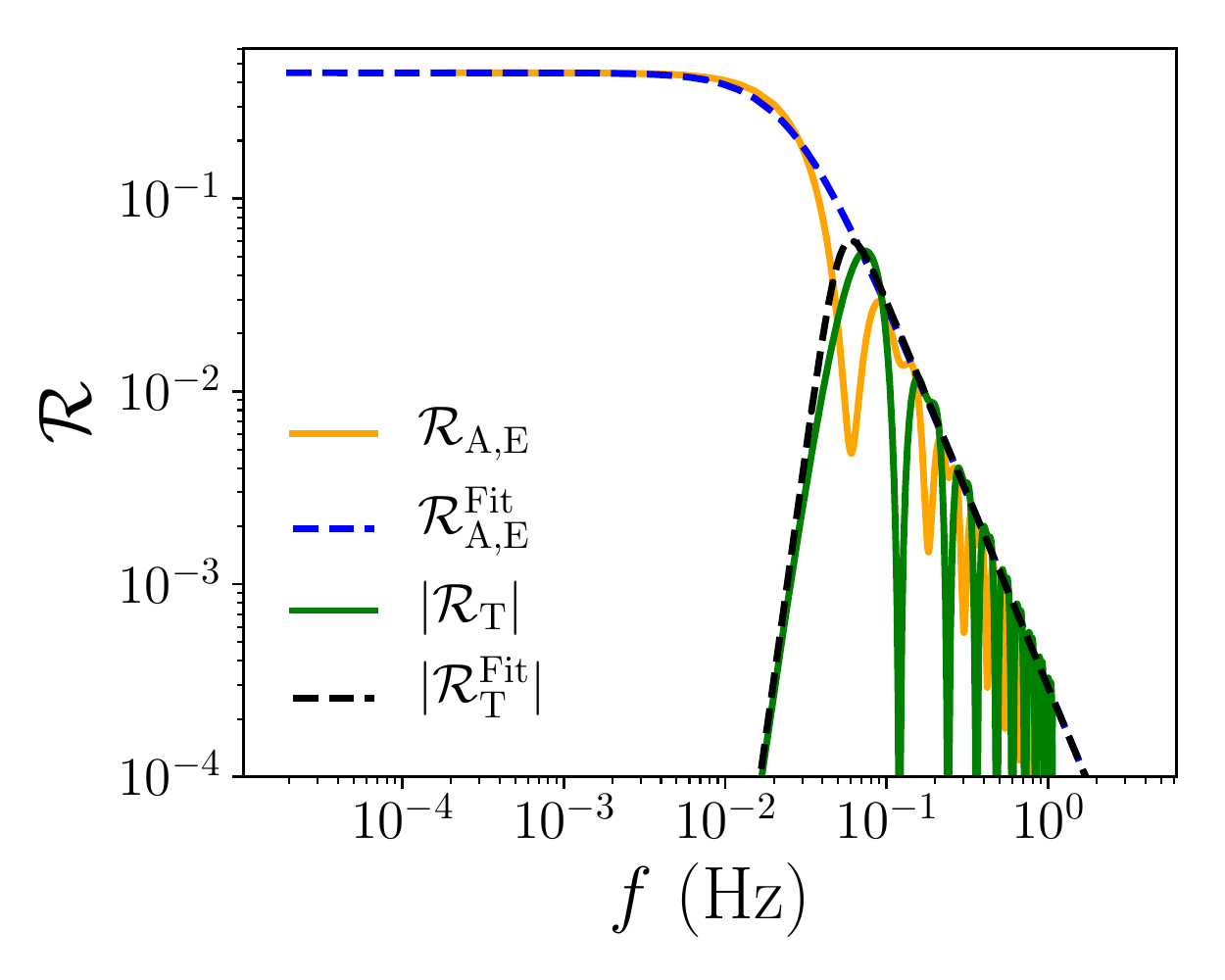}}
\caption{The LISA response functions ${\cal R}_{A,E}$ and ${\cal R}_T$ along with the fitting functions are shown as function of frequency.}
\label{fig:response}
\end{center}
\end{figure}

The A and E response functions are constant for $f\ll f_*$ and scale as $f^{-2}$ for $f>f_*$; the T response function goes as $f^6$ for $f \ll f_*$ and as $f^{-2}$ for $f>f_*$, as shown in Fig.~\ref{fig:response}. We find that an approximate fit for these response functions is given by
\begin{eqnarray}
  \mathcal{R}_{{\rm A,E}}^{\rm Fit} &\simeq& \frac{9}{20}|W|^2\left[1+\left(\frac{f}{ 4f_*/3}\right)^2\right]^{-1},\label{eq:RApprox}\\
  \mathcal{R}_{\rm T}^{\rm Fit} &\simeq& \frac{1}{4032} \left(\frac{f}{f_*}\right)^6|W|^2\left(1+\frac{5}{16128}\left[\frac{f}{ f_*}\right]^8\right)^{-1}.\label{eq:RApprox2}
\end{eqnarray}
The full response functions and the fits are shown in Fig.~\ref{fig:response}.
It is clear that the T mode is much less sensitive than the A and E modes. 

\section{The optimal statistic for a stochastic background}
\label{sec:optimal1}

Here we describe the procedure to obtain the optimal statistic for assessing sensitivity to a SGWB \cite{Allen:1997ad,Maggiore:1999vm,Romano:2016dpx}. At issue is how best to distinguish the signal from the noise. As discussed in Ref.~\cite{Hogan:2001jn}, the TDI T signal (also called the Sagnac signal), is much less sensitive to the SGWB at lower frequencies than the A and E signals. Because of this we can use the T signal to partially remove the instrumental noise from A and E. For simplicity we will assume that the T mode allows us to completely characterize the instrumental noise associated with the $\Phi_{A,E}$ modes, $N(t-t')$.  We will also ignore any effects due to the motion of the spacecraft, since these are expected to be negligible. This means that for a stationary SGWB the LISA response is also stationary \cite{Allen:1996gp,Cornish:2001bb}. 

Our assumption that the SGWB is stationary allows us to write down an optimal statistic of the form \cite{Allen:1996gp}
\begin{equation}
    \hat{\mathcal{C}} = \int_{-T/2}^{T/2}\int_{-T/2}^{T/2}\sum_{i={\rm A,E}} \left[\Phi_i(t) \Phi_i(t')-\frac{1}{2} N_i(t-t')\right] Q_i(t-t') dt dt',\label{eq:optstat}
\end{equation}
where $Q(t-t')$ is a weight that is chosen so as to maximize the signal-to-noise ratio (SNR) of this statistic.  Note that in order to obtain an unbiased statistic with respect to the SGWB we must subtract the instrumental noise. This might seem unnecessary since \emph{if} we had access to two signals whose correlation had zero response to the instrumental noise and non-zero response to the SGWB we could form an unbiased SGWB statistic without subtracting the instrumental noise. However, as noted after Eq.~(\ref{eq:noise1}), by diagonalizing the covariance, the cross-correlation between the TDI variables has zero response \emph{to both the instrumental noise and the SGWB}. This leaves us with using the autocorrelation between A and E while using the T signal to estimate and subtract the instrumental noise.   The expectation value of this statistic is given by
\begin{eqnarray}
    \mu \equiv \langle \hat{\mathcal{C}}\rangle &=& \frac{1}{2}\sum_{i={\rm A,E}}\int_{-T/2}^{T/2} \mathcal{S}_i(t-t')Q_i(t-t') dt dt',\\
    &=& \frac{1}{2}\sum_{i={\rm A,E}}\int_{-\infty}^\infty \delta^2_T(f_1+f_2){\mathcal{S}}_i(f_1){Q}^*_i(f_2)df_1 df_2,\label{eq:mu}
\end{eqnarray}
where $\delta_T(f_1+f_2) = T {\rm sinc}[(f_1+f_2) \pi T]$ is the finite-time approximation to the Dirac delta function. Note that the actual measurements will be discrete in time so that in order to write the Fourier transform we have assumed that the sampling rate is larger than the frequency support for the signal and weight. For a fixed $f_1$ when $|f_2|\gg |f_1|$ the finite-time delta function scales as $1/ (f_2)^2$. In addition to this the width of the finite-time delta function is $1/T$. As long as the filter function is smooth on scales of order $1/T$ and grows slower than $1/|f_2|$ we have 
\begin{equation}
    \delta^2_T(f_1+f_2) \simeq T \delta_D(f_1+f_2)
    \label{eqn:delta}
\end{equation} 
so that 
\begin{equation}
    \mu \simeq \frac{T}{2} \sum_{i={\rm A,E}} \int_{-\infty}^{\infty} {\mathcal{S}}_i(f)  {Q}_i(f) df.
\end{equation}
We will see that these conditions on the filter function are satisfied for a wide range of power-law SGWBs towards the end of this Section. 

The square of the estimator is given by 
\begin{eqnarray}
    \hat{\mathcal{C}}^2 &=& \int_{-T/2}^{T/2}\sum_{i={\rm A,E}} \sum_{j={\rm A,E}} \left[\Phi_i(t_1) \Phi_i(t_1')-\frac{1}{2}{N_i}(t_1-t_1')\right] \left[\Phi_j(t_2) \Phi_j(t_2')-\frac{1}{2}{N_j}(t_2-t_2')\right] \nonumber\\ &\times&Q_i(t_1-t_1')Q_j(t_2-t_2') dt_1 dt_1' dt_2 dt_2'.
\end{eqnarray}
To evaluate the expectation value of the square of the estimator, we must compute 
\begin{eqnarray}
&&   \bigg \langle \left[\Phi_i(t_1) \Phi_i(t_1')-\frac{1}{2}{N_i}(t_1-t_1')\right] \left[\Phi_j(t_2) \Phi_j(t_2')-\frac{1}{2}{N_j}(t_2-t_2')\right] \bigg \rangle \\
&=& \langle \Phi_i(t_1) \Phi_i(t_1')\Phi_j(t_2) \Phi_j(t_2')\rangle + \frac{1}{4} {N_i}(t_1-t_1'){N_j}(t_2-t_2') - \frac{1}{2}\langle \Phi_i(t_1) \Phi_i(t_1')\rangle{N_j}(t_2-t_2') - \frac{1}{2}\langle \Phi_j(t_2) \Phi_j(t_2')\rangle{N_i}(t_1-t_1').\nonumber
\end{eqnarray}
Computing each term separately, under the hypothesis that no signal has yet been detected, we have 
\begin{eqnarray}
\langle \Phi_i(t_1) \Phi_i(t_1')  \rangle & = & \frac{1}{2}N_i(t_1-t_1'),\\
 \langle \Phi_i(t_1) \Phi_i(t_1')\Phi_j(t_2) \Phi_j(t_2')\rangle & = & \frac{1}{4}\left(N_i(t_1-t_1') N_j(t_2-t_2')+ \delta_{ij}\left[N_i(t_1-t_2)N_j(t_1'-t_2') + N_i(t_1-t_2') N_j(t_1'-t_2)\right]\right),
\end{eqnarray}
where we use the fact that both A and E modes have the same noise power spectra. Combining all of these terms and summing over $i$ and $j$ we then find
\begin{eqnarray}
\sigma^2 = \langle \hat{\mathcal{C}}^2 \rangle &=& \frac{1}{2}\int_{-T/2}^{T/2} \left[N(t_1-t_2)N(t_1'-t_2') + N(t_1-t_2') N(t_1'-t_2)\right] Q(t_1-t_1') Q(t_2-t_2')dt_1 dt_1' dt_2 dt'_2,\\
&=&\frac{1}{2}\int_{-\infty}^{\infty}df_1 df_2 df_3 df_4 \delta_T(f_2-f_3)\delta_T(f_1+f_3)N(f_1)N(f_2)Q(f_3)Q(f_4)\nonumber \\ &\times&\left[\delta_T(f_1+f_4)\delta_T(f_2-f_4) +\delta_T(f_1-f_4)\delta_T(f_2+f_4)\right],\\
&\simeq&  T\int_{-\infty}^{\infty}df N^2(f)Q^2(f).
\end{eqnarray}
The SNR of this measurement is then given by 
\begin{equation}
{\rm SNR}= \frac{\mu}{\sigma} \simeq \sqrt{\frac{T}{2}} \frac{\sum_{i={\rm A,E}}\int_{-\infty}^\infty df \mathcal{S}_i(f) Q_i(f)}{\sqrt{\sum_{i={\rm A,E}}\int^{\infty}_{-\infty} df N_i^2(f)  Q_i^2(f)}}.
\end{equation}
Our retention of the sum over A, E is a formality, since the signal and noise is the same for the two detector eigenmodes. Hence, we can write
\begin{equation}
{\rm SNR}=   \sqrt{T} \frac{ \int_{-\infty}^\infty df \mathcal{S}_{\rm A}(f) Q_{\rm A}(f)}{\sqrt{\int^{\infty}_{-\infty} df N_{\rm A}^2(f)  Q_{\rm A}^2(f)}}.
\end{equation}
To determine what filter function $Q_{\rm A}(f)$ will maximize the SNR, we introduce a noise-weighted inner product 
\begin{equation}
(A,B) \equiv \int_{-\infty}^{\infty} df  A(f) B(f) N_{\rm A}^2(f).
\end{equation}
With this the SNR can be written as 
\begin{equation}
{\rm SNR} = \sqrt{T}\frac{\left(\mathcal{S}_{\rm A}/N_{\rm A}^2, Q_{\rm A}\right)}{\sqrt{(Q_{\rm A},Q_{\rm A})}}.
\end{equation}
It is clear that the SNR will be maximized if $Q_{\rm A}(f) = \lambda \mathcal{S}_{\rm A}(f)/N_{\rm A}^2(f)$, where $\lambda$ is some normalization.  With this choice, the optimal SNR is given by 
 \begin{equation}
{\rm SNR} = \left[T  \int_{-\infty}^\infty df \frac{\mathcal{S}_{\rm A}^2(f)}{N_{\rm A}^2(f)}\right]^{1/2}
= \left[T\sum_{i={\rm A,E}} \int_{0}^\infty df \frac{\mathcal{S}_i^2(f)}{N_i^2(f)}\right]^{1/2},
\label{eq:maxSNR}
\end{equation}
where in this last equality we have divided by two to rewrite the integrand as a sum over A and E modes, and multiplied by two in changing the range of integration.

As we will see next, at small frequencies $Q\propto f^8 S(f)$ due to the acceleration noise, and at large frequencies $Q\propto f^{-2} S(f)$ due to the frequency of the instrument response. Our ability to approximate the finite-time delta function as a Dirac delta function relies on the filter growing slower than $1/|f|$. This means that our SNR expression will be correct as long as we have $-5 < |\partial \ln S(f)/\partial \ln f|$ at $f \lesssim f_*$ and $ |\partial \ln S(f)/\partial \ln f|  < 3 $ at $f \gtrsim f_*$. If the SGWB has a power-law index outside of this range then we cannot approximate the finite-time delta function as a Dirac delta function and the SNR of this statistic will take a different form. 

\section{LISA Noise Model}
\label{sec:LISAnoise}

Expressions for the expected noise power spectra are given in the LISA Science Requirements Document \cite{LISASciRD}. The dominant sources of noise in this idealized treatment are due to acceleration noise and optical path-length fluctuations, with rms amplitudes
\begin{equation}
    \sqrt{\left(\delta a\right)^2} = 3 \times 10^{-15}\, {\rm m/s^2}, \, \qquad
    \sqrt{\left(\delta x\right)^2} = 1.5 \times 10^{-11}\, {\rm m}.
\end{equation}
The acceleration and optical metrology noise are given by
\begin{equation}
S_a = \frac{S_{I}}{4(2 \pi f)^4}, 
\qquad
S_s = S_{II}, \label{eq:accel_met_noise}
\end{equation} 
where
\begin{eqnarray}
    S_{I} &=& 4 \left(\sqrt{\left(\delta a\right)^2}/L\right)^2 (1 + (f_1/f)^2)\, {\rm Hz}^{-1}= 5.76 \times 10^{-48}(1 + (f_1/f)^2){\rm s}^{-4}\, {\rm Hz}^{-1}
    \label{eqn:SI}\\ \cr
    S_{II} &=& \left(\sqrt{\left(\delta x\right)^2}/L\right)^2 {\rm Hz}^{-1} = 3.6 \times 10^{-41} {\rm Hz}^{-1}
    \label{eqn:SII}
\end{eqnarray}
with $L=2.5\times 10^9$~m, $f_1 = 0.4$~mHz. 
These noise spectra contribute to the interferometer noise (see, e.g., Ref.~\cite{Cornish:2001bb})
\begin{eqnarray}
N_1 &=& \Big[ 4 S_s(f) + 8[1 +\cos^2(f/f_*)] S_a(f)\Big] |W|^2,\label{eq:noiseN1} \\
N_2 &=& -[2 S_s(f) + 8 S_a(f) ] \cos(f/f_*) |W|^2.\label{eq:noiseN2}
\end{eqnarray}
The noise of the A and E signals is given by
\begin{eqnarray}
    N_{\rm A,E} = N_1-N_2 &=& \left((4+2\cos(f/f_*))S_s + 8(1 +\cos(f/f_*)+\cos^2(f/f_*))S_a\right)|W|^2 \label{eq:Nexact} \\
    &\simeq & (6 S_s + 24 S_a)|W|^2. \label{eq:NApprox}
\end{eqnarray}
where the latter expression is obtained under a low-frequency approximation, $\cos(f/f_*) \simeq 1$, which provides a good fit to the exact noise curve without the high frequency wiggles.
We use these expressions for the noise in Eq.~(\ref{eq:maxSNR}). 

With these results in hand, we can determine the inverse noise-weighted response to the variance in the SGWB intensity or spectral density
\begin{equation}
\Sigma_{I} = \left[  \left(\frac{{\cal R}_{\rm A}  } {N_{\rm A}} \right)^2+\left(\frac{{\cal R}_{\rm E}  } {N_{\rm E}} \right)^2 \right]^{-1/2}, \qquad
\Sigma_\Omega = \Sigma_I \frac{4 \pi^2 f^3}{3 H_0^2}, \qquad
\Omega_{GW} = I \frac{4 \pi^2 f^3}{3 H_0^2}.
\label{eqn:sigmas}
\end{equation}
In terms of these new variables, the SNR is
\begin{equation}
\boxed{
    {\rm SNR}^2 = T \int_0^\infty df \, {I^2}/{\Sigma_I^2} = T \int_0^\infty df \, {\Omega_{GW}^2}/{\Sigma_\Omega^2}.}\label{eqn:altSNR}
\end{equation}
We repeat for the impatient reader that $T$ is the time of observation, e.g.~$4$ years.
We use Eqs.~(\ref{eqn:sigmas}-\ref{eqn:altSNR}) to evaluate the signal-to-noise ratio of LISA for a given SGWB.

The sensitivity to the intensity is therefore
\begin{eqnarray}
    \Sigma_I &=& \frac{1}{\sqrt{2}}\frac{N_{\rm A}}{{\cal R}_{\rm A}} \label{eqn:siexact}\\
    &\simeq& \frac{1}{\sqrt{2}} \frac{6 S_s + 24 S_a}{\frac{9}{20}\left[1+\left(\frac{f}{4f_*/3}\right)^2\right]^{-1}}.  \label{eqn:siapprox}
\end{eqnarray}
We refer to Eq.~(\ref{eqn:siexact}), using Eq.~(\ref{eq:Nexact}) for the noise and the exact expression Eq.~(\ref{eqn:rint}) for the response, as our exact, numerical result. The second expression, Eq.~(\ref{eqn:siapprox}), gives our approximation which uses the low-frequency approximation in Eq.~(\ref{eq:NApprox}) for the noise and the fitting function Eq.~(\ref{eq:RApprox}) for the response,
whereby
\begin{equation}
    \boxed{\Sigma_I \simeq \sqrt{2} \frac{20}{3} \left[ \frac{S_I(f)}{(2 \pi f)^4} + S_{II}(f) \right]\left[1+\left(\frac{f}{ 4f_*/3}\right)^2\right].}\label{eqn:box2}
\end{equation}
We note that the factors of $W$ for the extra TDI paths cancel exactly from both numerator and denominator in our idealized treatment. The boxed equations, with noise spectra given in Eqs.~(\ref{eqn:SI}-\ref{eqn:SII}), are sufficient to specify the signal-to-noise ratio of LISA for a given SGWB.

\section{Polarization- and sky-averaged LISA sensitivity for short-lived `multi-tonal' deterministic point sources}
\label{sec:skyavg}

The optimal SNR for a short-lived multi-tonal point source (i.e., a deterministic source that evolves through the LISA band in a short time compared to the mission lifetime) takes a slightly different form. Examples of such sources are a spinning neutron star or a binary merger. This is a standard calculation~\cite{Moore:2014lga}, although the detector response is not usually included for reasons of generality. We start by identifying the signal ${\cal S}$ as the interferometer phase $\Phi$ convolved with a filter, $Q$
\begin{equation}
   \hat {\cal C} = \sum_{i={\rm A,\,E}}\int_{-T/2}^{T/2} dt \, \Phi_i(t) Q_i(t)
    = \sum_{i={\rm A,\,E}} \int_{-\infty}^\infty df_1 df_2 \delta_T(f_1+f_2) \Phi_i(f_1) Q_i(f_2).
\end{equation}
The expectation value of this statistic is then given by 
\begin{equation}
  \mu \equiv  \langle \hat {\cal C} \rangle = \sum_{i={\rm A,\,E}} \int_{-\infty}^\infty df_1 df_2 \delta_T(f_1+f_2) \Delta \varphi_i(f_1)  Q_i(f_2).
  \label{eqn:mucont}
\end{equation}
If the source has a broad frequency dependence then the finite-time delta function approximates a Dirac delta; in other words, the source passes through the LISA band in a time much shorter than the observation. For the following discussion we will focus attention on those sources with a Fourier transform that is smooth on frequencies around and below $1/T$ so that we can write the signal as
\begin{equation}
   \mu  \simeq  \sum_{i={\rm A,\,E}} \int_{-\infty}^\infty df   \Delta \varphi_i(f) Q^*_i(f).
\end{equation}
The squared noise is the mean of the square of this same convolution in the absence of signal,
\begin{eqnarray}
    \sigma^2 &=& \sum_{i={\rm A,\,E}}\int dt \, dt'  Q_i(t)  Q_i(t') \, \langle \Phi(t)\Phi(t')\rangle,\\
    &=& \sum_{i={\rm A,\,E}}\int dt \, dt'  Q_i(t)  Q_i(t') \, \frac{1}{2} N_i(t-t'),\\
    &\simeq& \sum_{i={\rm A,\,E}} \int_{-\infty}^\infty df \, |Q_i(f)|^2 \, \frac{1}{2}N^*_i(f),
    \label{eqn:sig2cont}
\end{eqnarray}
where in the last line we again approximated the finite-time delta function as a Dirac delta function. 
Here we introduce the inner product
\begin{equation}
    (A,B) \equiv \int_{-\infty}^\infty df \, \frac{A(f) B(f)}{\frac{1}{2} N_i(f)}
\end{equation}
to facilitate writing the SNR as
\begin{equation}
    {\rm SNR}^2 = \frac{\mu^2}{\sigma^2} = \frac{\left(\sum_{i={\rm A,\,E}}(\Delta \varphi_i,\frac{1}{2}N_i Q_i) \right)^2}{\sum_{i={\rm A,\,E}}(\frac{1}{2}N_i Q_i,\frac{1}{2} N_i Q_i)}.
\end{equation}
This ratio is clearly maximized by the filter $Q_i = \lambda \Delta \varphi_i/\frac{1}{2} N_i$, where $\lambda$ is some normalization, whereupon
\begin{equation}
    {\rm SNR}^2 = \sum_{i={\rm A,\,E}} \left( \Delta \varphi_i, \Delta \varphi_i \right) = \sum_{i={\rm A,\,E}}\int_{-\infty}^\infty df \,\frac{|\Delta \varphi_i(f)|^2}{\frac{1}{2} N_i(f)}.
    \label{eqn:SNR2cont}
\end{equation}
If we were to rename the phase as $``\tilde h" = \Delta\varphi_i$, then we would recover the appearance of a standard formula
\begin{equation}
    {\rm SNR}^2 = \sum_{i={\rm A,\,E}} 4\int_{0}^\infty df \,\frac{|\tilde h(f)|^2}{N_i(f)}.\label{eq:std_SNR}
\end{equation}
(e.g. Eq.~16 of Ref.~\cite{Moore:2014lga}). However, the signal is the interferometer phase, so to take into account the response of the detector, we adapt Eqs.~(\ref{eqn:rdefn}) and (\ref{eqn:rint}) and we have 
\begin{eqnarray}
{\rm SNR}^2 &=&\sum_{i={\rm A,\,E}}4\int_{0}^\infty df \,\frac{|\Delta \varphi_i(f)|^2}{ N_i(f)}, \\
    |\Delta \varphi_i(f,\hat n_s)|^2 &=&  \big|\sum_P {h}_P(f) F^P_{i}(\hat n_s,f;t)\big|^2,\label{eq:phase_pt_source}
\end{eqnarray}
where $\hat n_s$ is a unit vector that points to the source. To calculate the SNR due to a specific object, this is the expression to use. Note that the direction of this vector will change in time due to the motion of LISA relative to the source location. However, we are going to proceed under the assumption that the instrumental response varies on the time-scale of days to weeks due to the orbital motion of the spacecraft whereas the signal oscillates on seconds to hour-long time-scales.

We proceed for the case of an average source, like a radiating binary. We assume that the source is in a plane with a normal that points in the $\hat u$ direction (for a binary this is the orbital angular momentum). This axis defines a coordinate system in which we can establish a natural basis on the sky 
\begin{equation}
    \hat e'_x \equiv \frac{\hat n_s \times \hat u}{|\hat n_s \times \hat u|},\ \hat e'_y \equiv -\frac{\hat n_s \times \hat e'_x }{|\hat n_s \times \hat e'_x |}.\label{eq:nat_basis}
\end{equation}
From this we can define a set of GW tensors in the usual way:
\begin{eqnarray}
    e^{'+}_{ab}(\hat n_s, \hat u) &=& (\hat e'_x)_a (\hat e'_x)_b - (\hat e'_y)_a(\hat e'_y)_b,\\
    e^{'\times}_{ab}(\hat n_s, \hat u) &=& (\hat e'_x)_a(\hat e'_y)_b + (\hat e'_y)_a(\hat e'_x)_b.
\end{eqnarray}
This set of polarization tensors are related to the detector-defined polarization tensors [see Eqs.~(\ref{eq:epab})-(\ref{eq:ecab})] through a polarization matrix \begin{equation}
    R^A_B(\psi) \equiv \left(\begin{array}{cc}\cos 2 \psi & \sin 2 \psi \\-\sin 2\psi & \cos 2 \psi\end{array}\right),
\end{equation}
where $e^A = R^A_B(\psi) e'^{B}$. Defining $\nu \equiv \hat n_s \cdot \hat u = \cos \theta_u$ the gravitational wave received at the detector can be written 
\begin{equation}
   h_{ab}(f) = A(f)\left[ g_+(\nu) R^+_A(\psi) e'^{A}_{ab}(\theta_u,\phi_u)+ g_\times(\nu) R^\times_A(\psi) e'^{A}_{ab}(\theta_u,\phi_u)\right].
   \label{eqn:habwaveform}
\end{equation}
The most agnostic assumption is that we do not have any prior information on the direction of $\hat u$. In this case we can average over the orientation $\hat u$ as well as the direction $\hat n_s$ of the source on the sky, in which case we take
\begin{eqnarray}
    |\Delta\varphi_i(f)|^2 &\rightarrow& \frac{1}{4\pi}\int d\Omega_u
    \frac{1}{4\pi}\int d\Omega_{n_s}\bigg|\sum_P {h}_P(f) F^P_{i}(\hat n_s,f;t)\bigg|^2, \label{eqn:averaging1}
    \\
    &=& \frac{1}{2}\bar{h}^2(f) {\cal R}_i(f),\label{eqn:averaging2}
    \\
    \bar{h}^2(f) &\equiv& \frac{A^2(f)}{2}\int d\theta_u \sin(\theta_u)\left[g^2_+(\theta_u) +g^2_\times(\theta_u)\right].
    \label{eqn:hwaveform}
\end{eqnarray}
Averaging over polarization is implicit in Eqs.~(\ref{eqn:averaging1}-\ref{eqn:averaging2}) so that
the polarization- and sky-averaged SNR is finally
\begin{equation}
     {\rm SNR}^2 =\sum_{i={\rm A,\,E}} 4\int_{0}^\infty df \,\frac{\mathcal{R}_i(f)}{N_i(f)}\frac{1}{2}\bar{h}^2(f)=
     \int_0^\infty df \, \frac{\bar{h}^2(f)}{\Sigma_{h}(f)}.
     \label{eqn:SNRdet}
\end{equation}
By collecting terms in the middle expression above, we define the noise power spectral density $\Sigma_h$ for short-lived sources in terms of the detector noise $N$ and response ${\cal R}$,
\begin{equation}
        \Sigma_h = \left( 2 \sum_{i={\rm A,\,E}} \frac{\mathcal{R}_i(f)}{N_i(f)} \right)^{-1}. \label{eq:exacth}
\end{equation}
The contribution of the T mode can be included by an obvious generalization of the above expression.
For the exact, numerical sensitivity we use Eqs.~(\ref{eq:Nexact}), (\ref{eqn:rint}) for the noise and response. For the low-frequency approximation, we use Eqs.~(\ref{eq:NApprox}), (\ref{eq:RApprox}). Hence, the signal-to-noise ratio is given by
\begin{equation}
     \boxed{{\rm SNR}^2 = \int_0^\infty df \, \frac{\bar{h}^2(f)}{\Sigma_{h}(f)} }
    \label{eqn:LISASNR}
\end{equation}
with
\begin{equation}
\boxed{\Sigma_{h} \simeq \frac{1}{2} \, \frac{20}{3} \, \left[ \frac{S_I(f)}{(2 \pi f)^4} + S_{II}(f) \right] R(f)}
\label{eqn:scird}
\end{equation}
where $R(f) = 1 + (f/f_2)^2$ and $f_2 = 25$~mHz (note that with $L=2.5 \times 10^9$ m we have $4f_*/3 = 25.4$ mHz). This SNR is useful for cases in which the waveforms are known. Eqs.~(\ref{eqn:LISASNR}), (\ref{eqn:scird}) match the expression for the SNR given in the LISA Science Requirements Document \cite{LISASciRD}.

The range of integration in the SNR for burst sources is set by considerations of the noise properties of the instrument (i.e., $f_{min}\simeq 10^{-5}$ Hz and $f_{\rm max} \simeq 1$ Hz). However, we can generalize this expression for the SNR to cover continuous wave sources which sit for an extended period of time within the LISA band by taking $f_{min} = {\rm max}(10^{-5}~{\rm Hz},f_{obs})$ and $f_{obs}$ is set by the time the object has been observed. Likewise, $f_{max} = {\rm min}(1~{\rm Hz}, f_{m})$ and $f_m$ is an upper frequency based on the source, like the ISCO for a binary merger. See Ref.~\cite{Cutler:1994ys} for details. 

It may be surprising that the sensitivity for a continuous source differs from that of a SGWB. However, the optimal statistic is responding to an important difference between these two cases. Namely for a continuous source the signal is linear in wave amplitude, whereas for a SGWB the signal is the intensity, which is quadratic in wave amplitude. Since $I \propto h^2$, it follows that the uncertainty in the intensity is related to the uncertainty in the wave amplitude $\delta I \propto 2 h \delta h$. Consequently, the sensitivity to $I$ should be half of that to $h$, assuming that we restrict our simplistic argument to one TDI mode. Indeed, revisiting Eqs.~(\ref{eqn:box2}) and (\ref{eqn:scird}) for one TDI mode, $\Sigma_{I,1} = \sqrt{2}\Sigma_I$ and $\Sigma_{h,1} = 2 \Sigma_h$. Comparing these expressions, we find that $\Sigma_{I,1} = 2 \Sigma_{h,1}$, as expected. A similar behavior for the sensitivity of pulsar timing arrays to a SGWB and deterministic signal has recently been pointed out in Ref.~\cite{Hazboun:2019vhv}.

\section{Polarization- and sky-averaged LISA sensitivity for `monotone' point sources}
\label{sec:monochrom}

The case of a long-lived monotone point source (i.e., a source that remains fixed within the LISA band for period of time of longer than the mission lifetime), such as a binary that is many years before merger, can be treated in a similar fashion to the short-lived source. We consider a gravitational wave that oscillates like a cosine with fixed amplitude. Consequently, the interferometer phase is $\Delta\varphi(t) = \Delta\varphi_m \cos(2 \pi f_m t + \phi)$, where the subscript $m$ indicates the source is monotone. The Fourier transform of this waveform is
\begin{equation}
    \Delta\varphi(f) = \frac{1}{2}\Delta\varphi_m \left(\delta(f+f_m) + \delta(f-f_m)\right). \label{eq:monotone_phase}
\end{equation}
We can insert this expression into Eq.~(\ref{eqn:mucont}) to obtain
\begin{eqnarray}
 \mu &=& \sum_{i={\rm A,\,E}} \int_{-\infty}^\infty df_1 df_2 \delta_T(f_1+f_2) \Delta \varphi_i(f_1)  Q_i(f_2) \\
 &=& \sum_{i={\rm A,\,E}} \int_{-\infty}^\infty df \, \frac{1}{2}\Delta\varphi_{m,i} \left(\delta_T(f+f_m) + \delta_T(f-f_m)\right)  Q_i(f).
\end{eqnarray}
The noise is unchanged from Eq.~(\ref{eqn:sig2cont}), so we may proceed directly to Eq.~(\ref{eqn:SNR2cont}), whereby
\begin{eqnarray}
    {\rm SNR}^2 &=& \sum_{i={\rm A,\,E}}\int_{-\infty}^\infty df \,\frac{|\frac{1}{2}\Delta \varphi_{m,i}\left(\delta_T(f+f_m) + \delta_T(f-f_m)\right)|^2}{\frac{1}{2} N_i(f)} \\
    &=& \sum_{i={\rm A,\,E}}\frac{1}{2}|\Delta\varphi_{m,i}|^2 \int_{-\infty}^\infty df \,\frac{|\left(\delta_T(f+f_m) + \delta_T(f-f_m)\right)|^2}{N_i(f)}.
\end{eqnarray}
Since $N(f)$ is slowly varying in the region where the finite-time delta functions have support, we can replace $\delta_T^2(f \pm f_m) \simeq T \delta_D(f \pm f_m)$. The SNR becomes
\begin{equation}
    {\rm SNR}^2 = \sum_{i={\rm A,\,E}} T \frac{|\Delta\varphi_{m,i}|^2}{N(f_m)_i}.
\end{equation}
This result is valid for a general, long-lived monotone source.
The sky- and polarization averaging is the same as before, $|\Delta\varphi_{m,i}|^2 = \frac{1}{2} \bar h^2_m {\cal R}_i(f_m)$.
Our final result for the SNR is therefore
\begin{equation}
    \boxed{{\rm SNR}^2 = \frac{1}{4} T \frac{\bar h^2_m}{\Sigma_h(f_m)}}
    \label{eqn:SNRmono}
\end{equation}
where $T$ is the observation time and $\bar h_m$ is the dimensionless, polarization-averaged waveform amplitude.

We can make a connection between the SNR for deterministic long-lived monotone and short-lived multi-tonal point sources by considering a process through which an evolving source radiates in narrow frequency intervals $[f_n-\frac{1}{2}\Delta f,\, f_n + \frac{1}{2}\Delta f]$ for a sequence of times $T_n$. We can adapt Eq.~(\ref{eqn:SNRmono}) to describe such a sequence as
\begin{equation}
    {\rm SNR}^2 = \sum_n \frac{1}{4} T_n \frac{\bar h^2_{m,n}}{\Sigma_h(f_n)}.
\end{equation}
Likewise, we can adapt Eq.~(\ref{eqn:LISASNR}) to describe radiation into a sequence of frequency intervals, to obtain
\begin{equation}
    {\rm SNR}^2 = \sum_n\int_{f_n - \frac{1}{2}\Delta f}^{f_n + \frac{1}{2}\Delta f} df \, \frac{\bar{h}^2(f)}{\Sigma_{h}(f)} 
    = \sum_n\int_{t_n - \frac{1}{2}\Delta t}^{t_n + \frac{1}{2}\Delta t} dt \, \frac{df}{dt} \frac{\bar{h}^2(f)}{\Sigma_{h}(f)}
    = \sum_n T_n \frac{df}{dt} \frac{\bar{h}^2(f_n)}{\Sigma_{h}(f_n)}.
\end{equation}
We assume that the detector noise $\Sigma_h$ is very slowly varying across the frequency intervals. These two expressions for the SNR are equivalent, provided that
\begin{equation}
    \frac{1}{4} \bar h_{m,n}^2 = \frac{df}{dt} \bar h^2(f_n).
    \label{eqn:connection}
\end{equation}
The above equation is indeed valid. As shown in Ref.~\cite{Cutler:1994ys}, given a function $B(t) = A(t) \cos\phi(t)$, then under a pair of conditions that restrict the rate of change of $\phi$, and which are satisfied for both monotone and multi-tonal sources, the Fourier transform is $ B(f) = \frac{1}{2} A(t) \left(\frac{df}{dt}\right)^{-1/2}$. In our nomenclature, $\bar h(f_n) = \frac{1}{2} \bar h_{m,n} \left(\frac{df}{dt}\right)^{-1/2}$, so that Eq.~(\ref{eqn:connection}) is proved. Our results for monotone and multi-tonal sources are consistent.

\section{Worked Examples}
\label{sec:examples}

Here we will present worked examples of how to apply the SNR expressions to determine sensitivity to various sources of cosmological interest.

\subsection{LISA sensitivity curve to a SGWB}
\label{sec:integrated}

\begin{figure}[h]
\begin{center}
\resizebox{!}{8cm}{\includegraphics{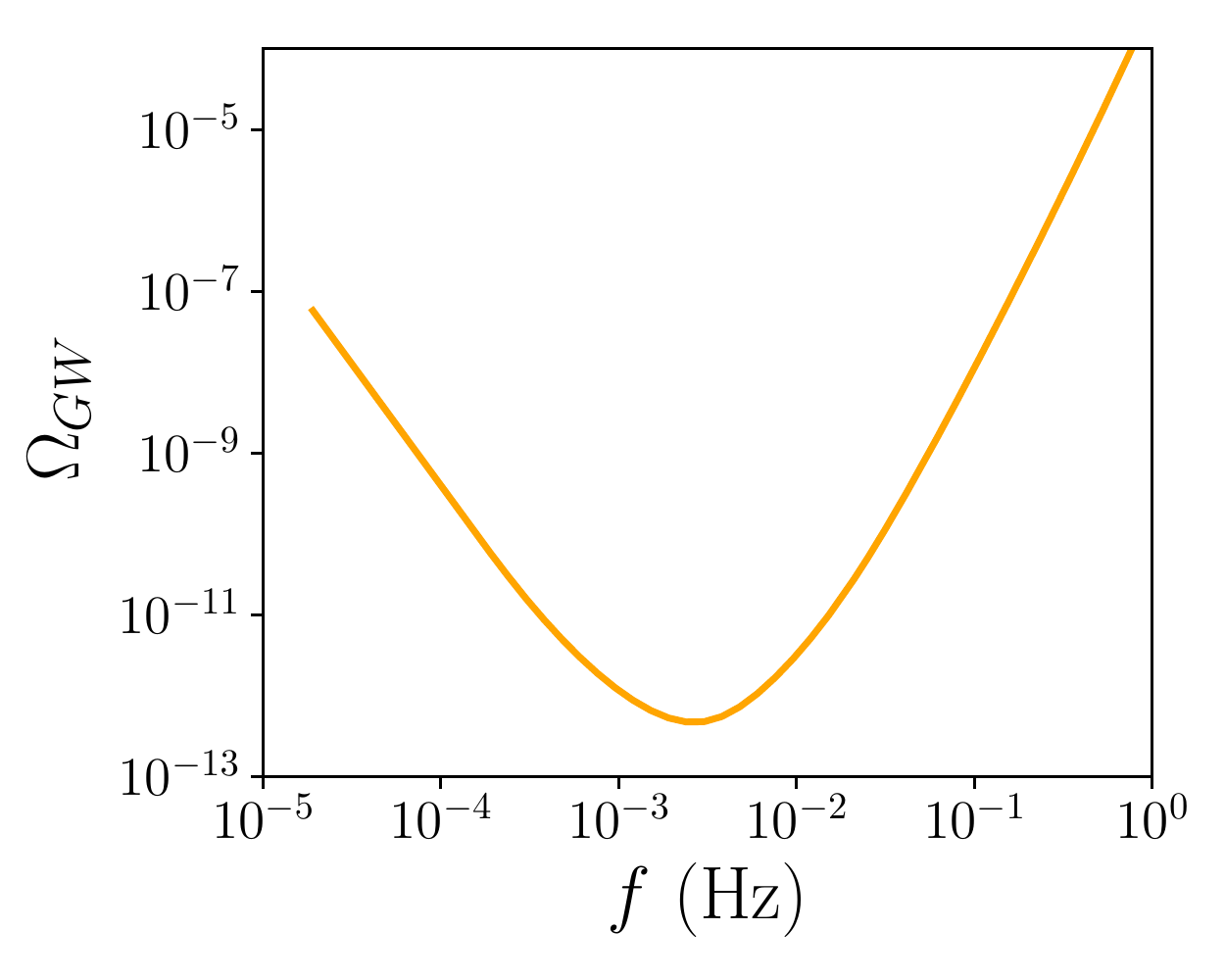}}
\caption{The integrated sensitivity of LISA to a power-law SGWB with SNR = 5 and observation time $T=4$~years. }
\label{fig:SGWBSens}
\end{center}
\end{figure}

We now compute the sensitivity of LISA to a SGWB. We require the signal to noise ratio as given in Eqs.~(\ref{eqn:sigmas}-\ref{eqn:altSNR}),
\begin{equation}
    {\rm SNR} = \left[T \int_{f_{\rm min}}^{f_{\rm max}} df \, \frac{\Omega_{\rm GW}^2 }{\Sigma_\Omega^2}\right]^{1/2},
\end{equation}
to exceed ${\rm SNR}=5$ for $T=4$~years observation.  A reasonable range of frequencies is $f_{min} = 0.1$~mHz and $f_{max}=0.1$~Hz. In the case of a flat, scale-invariant spectrum, assuming $H_0 = 67$~km/s/Mpc \cite{Aghanim:2018eyx}, we obtain $\Omega_{\rm GW} = 4.7 \times 10^{-13}$, or $\Omega_{\rm GW} h^2 = 2.1 \times 10^{-13}$. We stress that this includes both A and E TDI modes. This is an idealization, which we have made clear throughout, as we ignore foreground contamination, non-Gaussianities, data interruptions, and other systematics. For example, we expect that the scientific data set will be shorter, due to engineering tests and cuts in the data. The LISA Science Requirements document \cite{LISASciRD} projects that ``the duty cycle of usable science data at full (nominal) performance shall be greater than $75\%$,'' which we interpret to mean a $T \ge 3$~year data set. To adjust our forecast accordingly, $\Omega_{GW}$ is scaled upwards by a factor $\sqrt{4/3}$. Hence, we obtain $\Omega_{\rm GW} = 5.4 \times 10^{-13}$, or $\Omega_{\rm GW} h^2 = 2.4 \times 10^{-13}$.

In the case of a scale-free, power-law spectrum, we use the method of Thrane and Romano \cite{Thrane:2013oya} to draw the integrated sensitivity curve. The procedure is as follows.
\begin{enumerate}
\item We model the SGWB as a power-law so that $\Omega_{GW}(f) = \Omega_{GW,0} (f/f_0)^{n_{T}}$. \item For each value of the spectral index $n_T$ we determine the threshold $\Omega_{GW,0}$ that yields ${\rm SNR}=5$. 
\item We draw the maximum of the locus of curves consisting of $\Omega_{GW}(f)$ for each value of $n_T$ and $\Omega_{GW,0}$, at each frequency. This gives the integrated sensitivity curve.
\end{enumerate}
The integrated SGWB sensitivity curve using all TDI modes is shown in Fig.~\ref{fig:SGWBSens}. The curve shows that by combining information from the A and E modes, a four-year-long LISA mission can detect a SGWB as low as $\Omega_{\rm GW} \simeq 4.7 \times 10^{-13}$ with SNR = 5. Any power-law spectrum that intersects with the above curve is detectable at SNR=5.

In the case of a SGWB with any other shape within the sensitivity range, e.g.~a broken power-law, then drawing such a curve is only useful to guide expectations. The sensitivity is determined by evaluating the SNR on a case-by-case basis.

\subsection{Computing the Fisher matrix}\label{sec:Fisher}

We can also use a Fisher analysis to determine how well a gravitational wave observatory can infer parameters associated with the spectral density of a SGWB \cite{Tegmark:1996bz,Kuroyanagi:2018csn}. In order to perform this analysis we need to identify the data and then compute its covariance. Laser interferometers monitor the phase difference between light traveling along different paths, $\Phi_a (t_i) = \Delta \varphi_a(t_i) + n_a(t_i)$, where $\Delta \varphi_a(t_i)$ is due to gravitational waves and $ n_a(t_i)$ is the instrumental noise. We will denote the time interval between these measurements by $\Delta t = t_{i+1}-t_i$. 

We will divide up the total dataset of duration $T$ into time intervals of duration $1/f_l$, where $f_l$ is the highest frequency we are interested in (for LISA's nominal design $f_l = 0.1$ Hz; see Fig.~\ref{fig:SGWBSens}).  We imagine performing a Fourier analysis on each interval and will assume that different intervals are statistically independent (this is a better approximation the further separated the intervals get in time) \cite{Karnesis:2019mph,Caprini:2019pxz}. In this way we obtain $M\equiv f_lT$ quasi-independent measurements in each frequency bin $f_i$. The `data' is then given by each phase measurement, $\{\Phi^{(1)}_a(f_{i}),\Phi^{(2)}_a(f_{i}),\dots,\Phi^{(M)}_a(f_{i})\}$, for the two independent modes $a={\rm A,E}$. The mean of the data vanishes and its covariance is diagonal and is given by 
\begin{eqnarray}
{\bf C} &\equiv& \langle \Phi^{(p)}_a(f_i) \Phi^{(q)}_{b}(f_j)\rangle, \\
&=& \frac{1}{2}\left[\mathcal{S}_a(f_i) + N_a(f_i)\right] \delta_{ij} \delta_{pq}\delta_{ab}.
\end{eqnarray}
Assuming that the data is a realization of a Gaussian distribution, then the Fisher information matrix is given by \cite{Tegmark:1996bz}
\begin{eqnarray}
F_{\alpha \beta} &=& \frac{1}{2}{\rm Tr} \left[{\bf C}^{-1} \frac{\partial{\bf C}}{\partial \theta_\alpha}{\bf C}^{-1} \frac{\partial{\bf C}}{\partial \theta_\beta}\right],\\
&=& \frac{1}{2}M \sum_{a=A,E}\sum_i  \frac{\frac{\partial\mathcal{S}_a(f_i)}{\partial \theta_\alpha}\frac{\partial\mathcal{S}(f_i)}{\partial \theta_\beta}}{\left[N_a(f_i)+\mathcal{S}_a(f_i)\right]^{2}},\\
&\simeq& \frac{1}{2}T \sum_{a=A,E}\int_{f_l}^{f_h} \frac{\frac{\partial\mathcal{S}_a(f)}{\partial \theta_\alpha}\frac{\partial\mathcal{S}_a(f)}{\partial \theta_\beta}}{\left[N_a(f)+\mathcal{S}_a(f)\right]^{2}}df,
\end{eqnarray}
where we integrate to maximum frequency $f_h \simeq 1/\Delta t$, $\theta_a$ is a parameter used to model the SGWB, and we have assumed that the instrumental noise can be completely characterized by monitoring the Sagnac ($T$-mode) signal. Rewriting the signal in terms of $\Omega_{GW}$ \cite{Maggiore:1999vm} we have 
\begin{eqnarray}
 F_{\alpha \beta} = \frac{9 H_0^4}{32 \pi^4}T  \sum_{a=A,E}\int_{f_l}^{f_h} \frac{\frac{\partial\Omega_{GW}(f)}{\partial \theta_\alpha}\frac{\partial\Omega_{GW}(f)}{\partial \theta_\beta}\mathcal{R}_a^2(f)}{[N_a(f)+\frac{3 H_0^2}{4 \pi^2f^3}\Omega_{GW}(f)\mathcal{R}_a(f)]^2}\frac{df}{f^6}.\nonumber
 \end{eqnarray}
 The inverse of the Fisher matrix is the parameter covariance matrix giving us estimates for their uncertainties (see, e.g., Ref.~\cite{2009arXiv0906.4123C}).  

\subsection{A Binary Inspiral}\label{sec:binary}

We consider the sensitivity of LISA to a black hole binary inspiral far from merger. We describe such a system in terms of a waveform
\begin{eqnarray}
 \tilde h_{+} &=& A(f) \frac{1 + \cos^2\theta_u}{2} \cos\Psi \\
  \tilde h_{\times} &=& A(f) \cos\theta_u \sin\Psi
\end{eqnarray}
where $\theta_u$ describes the inclination of the orbit relative to our line of sight, and $\Psi$ is the phase. Considering Newtonian orbits, the leading-order contribution to the amplitude is
\begin{equation}
    A(f) = \sqrt{\frac{5}{24}}\frac{(G {\cal M}/c^3)^{5/6}}{\pi^{2/3} (D/c)}f^{-7/6},
\end{equation}
valid for frequencies far below the frequency at the innermost stable compact orbit, where ${\cal M}$ is the chirp mass and $D$ is the comoving distance. (See Refs.~\cite{Cutler:1994ys,Cornish:2018dyw} for more details.) Note that ${\cal M}$ and $f$ are in the source reference frame. Adapting Eqs.~(\ref{eqn:habwaveform}), (\ref{eqn:hwaveform}), we obtain
\begin{equation}
    \bar h^2(f) = \frac{4}{5}A^2(f) = \frac{(G {\cal M}/c^3)^{5/3}}{6 \pi^{4/3} (D/c)^2} f^{-7/3}.
    \label{eqn:hsource}
\end{equation}
We can apply this to Eqs.~(\ref{eqn:LISASNR}), (\ref{eqn:scird}) to evaluate the signal-to-noise ratio. If we assume that the binary is radiating when the detector turns on, and subsequently evolves out of the sensitivity window, then the range of frequency for integration of the SNR extends from the initial frequency up to the highest frequency detectable, which is about $1$~Hz.

\begin{figure}[h]
\begin{center}
\resizebox{!}{6cm}{\includegraphics{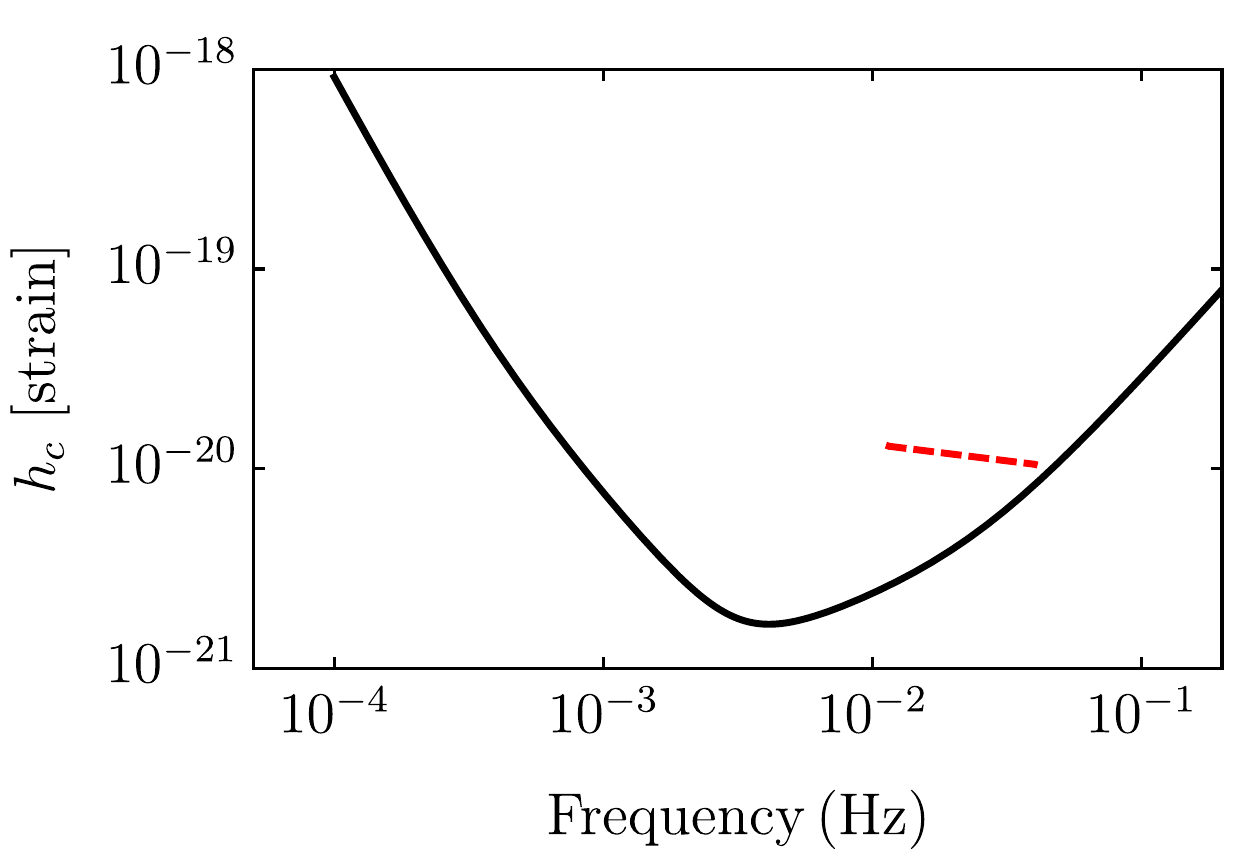}}
\caption{The characteristic strain $h_c$ due to a system identical to GW15094 \cite{Abbott:2016blz} that is radiating in the LISA band for 10 years prior to merger (dashed line), is shown relative to the strain sensitivity $2\sqrt{f \Sigma_h}$ (solid line)  calculated from Eq.~(\ref{eqn:scird})}.
\label{fig:bbh}
\end{center}
\end{figure}

As a classic example \cite{Sesana:2016ljz}, we consider a system identical to GW150914 \cite{Abbott:2016blz} that is radiating in the LISA band for several years prior to merger. We take $M_1 = 36 M_\odot$, $M_2=29 M_\odot$, and place the binary at $z=0.09$ in a standard cosmology with $\Omega_M=0.3$, $\Omega_\Lambda=0.7$, $h=0.7$ so that $D_L = 411.5$~Mpc. The chirp mass is ${\cal M} = \left((M_1 M_2)^3/(M_1+M_2)\right)^{1/5}\simeq 28 M_\odot$. We consider the situation that when LISA first spots the binary, it is radiating at a frequency $f_r$ in the binary rest frame, or $f_i = f_r/(1+z)$ in the reference frame of the observer. This predicts a time to merger in the binary frame
\begin{equation}
    t_{merge} = \frac{5}{256 \pi^{8/3}}\left(G {\cal M}/c^3\right)^{-5/3} f_r^{-8/3}
\end{equation}
which is $\Delta t_{obs}=(1+z) t_{merge}$ in the observer frame. For convenience, we imagine observing the binary in the LISA window 4 (10) years before merger, so that $\Delta t_{obs}\simeq 4\, (10)$~years implies $f_r= 0.018\, (0.013)$~Hz and $f_i=0.016 \, (0.012)$~Hz \cite{Sesana:2016ljz}. Using Eq.~(\ref{eqn:hsource}), we find that the characteristic strain is
\begin{equation}
    h_c = h_c(f_i) \left({f}/{f_i}\right)^{-1/6},\quad h_c(f_i) = \sqrt{\frac{2}{3}}\frac{(G {\cal M} (1+z)/c^3)^{5/6} f_i^{-1/6}}{\pi^{2/3} (D_L/c)}
\end{equation}
with $h_c(f_i) \simeq 1.2 \times 10^{-20}$ and $\bar h = h_c/(2 f)$, and $f$ is now in the reference frame of the observer. We can now use Eqs.~(\ref{eqn:LISASNR}) and (\ref{eqn:scird}), integrating from $f_i$ up to $f_{max} =1$~Hz, 
\begin{equation}
    {\rm SNR}^2 = \int_{f_i}^{f_{max}} df \frac{\bar h^2(f)}{\Sigma_h(f)} = \int_{f_i}^{f_{max}} d\ln f \frac{h_c^2(f)}{4 f \Sigma_h(f)}. 
    \label{eqn:GW150914snr}
\end{equation}
 Fig.~\ref{fig:bbh} shows the characteristic strain $h_c$ relative to the characteristic strain sensitivity, $\sqrt{4 f \Sigma_h}$. A good rule of thumb for assessing detectability is that the strain must lie half an order of magnitude above the strain sensitivity over an order of magnitude span in frequency. In our worked case, the result is ${\rm SNR}= 2.3\, (3.4)$ as shown in Table~\ref{tab:snr}, which is just at the threshold of detection. We have also verified that, given the frequency of GW150914, the T mode contributes negligibly to the SNR.

 \begin{table}[h]
     \centering
{
    \begin{tabular}{|l|l|l|} \hline
     \multicolumn{3}{|c|}{GW150914 Benchmarks} \\ [0.5ex] \hline
     Time to merger &   4 years &   10 years \\ \hline  
     ${\rm min}(f)|_{obs}$ & 0.016 Hz & 0.012 Hz \\ \hline
     ${\rm SNR}$, Eqs.~(\ref{eqn:LISASNR},\ref{eqn:scird}) &  $2.3$ & $3.4$ \\
     ${\rm SNR}$, Eqs.~(\ref{eq:exacth},\ref{eqn:LISASNR}) &  $2.7$ & $3.8$ \\ \hline
     \end{tabular}
     \caption{The SNR for LISA to observe a system identical to GW150914 under various conditions. The left (right) column shows the case that the binary is radiating for 4 (10) years before merger in the reference frame of the observer. The SNR is calculated first using the low-frequency approximation to the sensitivity, given in Eq.~(\ref{eqn:scird}). Second, the SNR is calculated using the exact, numerical results for the noise and response functions in Eq.~(\ref{eq:exacth}). All SNR values include two independent TDI modes.} 
     \label{tab:snr}
     }
 \end{table}

\subsection{A Monotone Binary}\label{sec:monoexample}

We consider the sensitivity of LISA to a binary system that is far from merger, which emits at essentially a single frequency. We describe such a system in terms of a waveform
\begin{eqnarray}
 \tilde h_{+} &=& A_m \frac{1 + \cos^2\theta_u}{2} \cos\Psi \\
  \tilde h_{\times} &=& A_m \cos\theta_u \sin\Psi
\end{eqnarray}
where $\theta_u$ describes the inclination of the orbit relative to our line of sight, and $\Psi$ is the phase. In this case, again considering Newtonian orbits, the leading contribution to the amplitude is
\begin{equation}
    A_m = 4 \frac{(G {\cal M})^{5/3} (\pi f_m)^{2/3}}{D c^4}.
\end{equation}
(See Ref.~\cite{Thorne:1987af} for more details.) Adapting Eqs.~(\ref{eqn:habwaveform}), (\ref{eqn:hwaveform}), we obtain $\bar h_m^2 = \frac{4}{5}A^2_m$, so
\begin{equation}
    \bar h_m^2 = \frac{64}{5} \frac{(G {\cal M})^{10/3} (\pi f_m)^{4/3}}{(D c^4)^2}.
\end{equation}
We can apply this to Eqs.~(\ref{eqn:scird}), (\ref{eqn:SNRmono}) to evaluate the signal-to-noise ratio.

As a timely example, we consider a system comparable to the recently discovered double white-dwarf binary ZTFJ1539 \cite{Burdge:2019hgl}. We use $M_1 = 0.6 M_\odot$ and $M_2 = 0.2 M_\odot$ so that ${\cal M}=0.3 M_\odot$, the orbital period is $P=415$~s so that $f_m = 4.8\times 10^{-3}$~Hz, and we place the object at a distance $2.3$~kpc, where the difference between physical and comoving distance is irrelevant. The time to merger is over $2\times 10^5$ years, and the radiating frequency evolves very slowly, changing by $10^{-6}$ per year. Hence, we are justified to treat this system as a monotone source. Using these numbers, the averaged strain amplitude is $\bar h_m = 1.8\times 10^{-22}$, and the noise power spectral density is $\Sigma_h(f_m) = 1.5 \times 10^{-40}~{\rm Hz}^{-1}$. For $T=1,\, 4$ years observation, the sky- and polarization-averaged SNR is  $40,\, 80$. All SNR values include two independent TDI modes. This source should be clearly observed by LISA \cite{Littenberg:2019mob}.

\section{Discussion}
\label{sec:discuss}

We have presented expressions for the optimal signal-to-noise ratio for LISA, in particular the power spectral density $\Sigma_\Omega$ for sensitivity to a SGWB and $\Sigma_h$ for the sky- and polarization-averaged sensitivity to a deterministic source. We have illustrated each with a worked example. We envision that these tools should enable a cosmologist to be able to assess the detectability of any new source of gravitational waves. These examples include benchmarks for easy comparison of methods. LISA should be able to observe a SGWB with $\Omega_{GW}h^2 = 2.1 \times 10^{-13}$ assuming $T=4$~years at ${\rm SNR}=5$. A binary black hole system that is identical to GW150914, radiating for 4 years prior to merger, would be marginally resolved with ${\rm SNR}=2.3$. A nearby double white-dwarf binary similar to ZTFJ1539 should be clearly detected with one year of observation, with ${\rm SNR}=40$. All three results are obtained through idealized calculations that ignore the presence of foregrounds and other systematic effects beyond the noise model.

Additional, independent sources of noise, beyond the instrumental noise modeled herein, can be included easily in the SNR expressions. For example, a noise spectral density representing a foreground of unresolved sources $\Delta N_i$ can be included by replacing $N_i \to N_i + \Delta N_i$ in Eqs.~(\ref{eq:maxSNR}) and (\ref{eqn:SNRdet}). Covariance or cross-correlation across different detectors is straightforward to calculate, but is beyond the scope of this article.

The tools we have presented may be naively extended to other space-borne gravitational wave observatories, such as TianQin \cite{Luo:2015ght} and the DECi-hertz Interferometer Gravitational wave Observatory (DECIGO). TianQin is a LISA-like constellation of three drag-free spacecraft, but orbiting the Earth with separation $L=\sqrt{3}\times 10^{8}$~m. The targeted acceleration noise and optical path-length fluctuation rms amplitudes are $\sqrt{(\delta a)^2} = 10^{-15}\,{\rm m}/{\rm s}^2$ and $\sqrt{(\delta x)^2} = 10^{-12}\,{\rm m}$. Assuming identical TDI modes as for LISA, then the equations in Secs.~\ref{sec:LISAnoise} and \ref{sec:skyavg} can be adapted (and rescaled, as for $f_1$) to obtain $\Sigma_\Omega$ for sensitivity to a SGWB and $\Sigma_h$ for polarization- and sky-averaged sensitivity to a deterministic source. DECIGO is another LISA-like constellation, but with arm length $L=1000$~km.

We provide a Mathematica notebook, available to download from our url, to facilitate easy computation. The notebook contains easy to use tools for SGWB studies. This includes a data table for $\Sigma_I$ using the calculated response functions, the analytic expression for $\Sigma_I$ to enable fast calculation of SNR for a SGWB, as well as a data table for the ${\rm SNR}=5$ integrated sensitivity curve shown in Fig.~\ref{fig:SGWBSens} for easy graphing. For studies of continuous sources, a data table and analytic expression for $\Sigma_h$ are included, as well as a data table for the strain sensitivity curve shown in Fig.~\ref{fig:bbh}.

At this moment, during this first stage of the age of gravitational wave astronomy, it is clearly necessary for the gravitational wave community to provide clear, simple, and accurate tools with which researchers in adjacent fields can estimate the sensitivity of current and future gravitational wave observatories to a variety of sources.  The current state of the literature on this topic can be at best 
described as confusing, with a proliferation of various notations and 
conventions; there are several popular online tools that produce 
sensitivity curves for LISA but which are significantly out of date, e.g., Refs.~\cite{shanesens,gwplotter}

In this paper we have provided a derivation of this sensitivity curve from start to finish along with specific examples of the SNR for LISA due to  a handful of standard sources. Our hope is that this paper provides a clear guide for any researcher looking to estimate the SNR for LISA due to any source and can provide the basic building blocks to assess the sensitivity of other space-based gravitational wave observatory designs.

\acknowledgments

TLS and RC thank Emanuele Berti, Chiara Caprini, Matthew Digman, Antoine Petiteau, Kaze Wong, and other members of the LISA consortium for patience and helpful discussions.

\vfill

\bibliography{LISASNR.bib}

\begin{appendix}
\renewcommand{\arraystretch}{1.4}
\begin{center}
\begin{table}[b]
\begin{tabular}{| c | c | c |}
\hline 
Symbol & Description (dimensions) & Defining equation\\
\hline \hline
$f$ & GW frequency &  \eqref{eq:Fourier}\\
$\hat{n}$ & unit vector along the GW direction of propagation & \eqref{eq:Fourier}\\
$h_{ab}(f, \hat{n})$ & Fourier transform of the GW strain (frequency$^{-1}$) & \eqref{eq:Fourier} \\
$e_{ab}^{+,\times}(\hat n)$ & linear gravitational wave polarization tensor & \eqref{eq:epab} \& \eqref{eq:ecab}\\
 $I(f)$ & total intensity of the SGWB (frequency$^{-1}$) & \eqref{eq:stokes}\\
 $\Omega_{\rm GW}$ & energy density of SGWB per log $f$ in units of the critical energy density& \eqref{eq:Omega_gw}\\
 $\Delta \varphi_{A_{BC}}(t)$ & laser phase accumulated at detector vertex due to GW& \eqref{eq:int_phase}\\
 $n_{A_{BC}}(t)$ & phase noise accumulated at detector vertex & \eqref{eq:int_phase}\\
 $F^P_{A_{BC}}(\hat n,f;t)$ & geometrical function describing gain of a detector vertex & \eqref{eq:gain1} \& \eqref{eq:gain2} \\
  $W$ & The `TDI' factor& \eqref{eq:gain2} \\
  $f_*$ & characteristic interferometer frequency & \eqref{eq:gain2} \\
  $L$ &  interferometer arm length& \eqref{eq:gain2} \\
 $\mathcal{R}_{A_{BC},X_{YZ}}(f;t,t')$ & SGWB intensity response& \eqref{eqn:rint} \\
 $N_{X={\rm A,E,T}}$ & TDI spectral noise response (frequency$^{-1}$)& \eqref{eq:tdinoise1} \& \eqref{eq:tdinoise2} \\
 $\mathcal{S}_{X={\rm A,E,T}}$ & TDI spectral signal response (frequency$^{-1}$)& \eqref{eq:tdisig1} \& \eqref{eq:tdisig2} \\
 $\mathcal{R}^{\rm fit}_{X={\rm A,E,T}}$ & fit to TDI intensity response & \eqref{eq:RApprox}  \& \eqref{eq:RApprox2} \\
 $T$ & lifetime of the mission & \eqref{eq:optstat} \\
 $\hat{\mathcal{C}}$ & optimal statistic & \eqref{eq:optstat} \\
 $Q(t-t')$ & optimal weight & \eqref{eq:optstat} \\
 $\delta_T(f_1+f_2)$ & finite-time delta function (frequency$^{-1}$) & \eqref{eq:mu} \\
 $S_a$ & acceleration noise spectral density (frequency$^{-1}$) & \eqref{eq:accel_met_noise} \\
  $S_s$ & optical metrology noise spectral density (frequency$^{-1}$) & \eqref{eq:accel_met_noise} \\
  $\Sigma_X$ & noise-weight response & \eqref{eqn:sigmas} \\
   $\hat n_s$ & unit vector pointing to a point-source & \eqref{eq:phase_pt_source} \\
   $\hat u$ & unit vector normal to point-source plane & \eqref{eq:nat_basis} \\
   $\nu$ & cosine of angle between $\hat n_s$ and $\hat u$ & \eqref{eqn:habwaveform} \\
   $A(f)$ & gravitational wave amplitude & \eqref{eqn:habwaveform} \\
    $\mathcal{M}$ & chirp mass & \eqref{eqn:habwaveform} \\
 \hline
\end{tabular}
\caption{Summary of the notation used in this paper} \label{tab:notation}
\end{table}
\end{center}
\end{appendix}

\end{document}